\documentclass[twocolumn]{aa}
\usepackage{graphics}
\begin{document}


\title{Color Effects Associated with the 1999 Microlensing Brightness Peaks
   in Gravitationally Lensed Quasar \object Q2237+0305}

\author{V.G.Vakulik \inst{1}
       \and
        R.E.Schild \inst{2}
       \and 
        V.N.Dudinov \inst{1}
       \and
        A.A.Minakov \inst{3}
       \and
        S.N.Nuritdinov \inst{4} 
       \and
        V.S.Tsvetkova \inst{3}
       \and
        A.P.Zheleznyak \inst{1}
       \and
        V.V.Konichek \inst{1}
       \and
       I.Ye.Sinelnikov \inst{1}
       \and 
       O.M.Burkhonov \inst{4}
       \and 
       B.P.Artamonov \inst{5}
       \and
       V.V.Bruevich \inst{5}}

\offprints{V.Vakulik}

\institute{Institute of Astronomy of Kharkov National University, Sumskaya
           35, 61022 Kharkov, Ukraine\\
           email: vakulik@astron.kharkov.ua
\and
           Center for Astrophysics, 60 Garden Street, Cambridge, MA
           02138, U.S.A.\\
           email: rschild@cfa.harvard.edu
\and
           Institute of Radio Astronomy of Nat.Ac.Sci. of Ukraine,
           Chervonoznamennaya 4, 61002 Kharkov, Ukraine\\
           email: minakov@ira.kharkov.ua
\and       
           Ulugh Beg Astronomical Institute of Ac.Sci. of Uzbekistan,
           Astronomicheskaya 33, 700052, Tashkent, Republic of Uzbekistan\\
           email: nurit@astrin.uzsci.net
\and
           Sternberg Astronomical Institute, Universitetski Ave. 13, 
           119899 Moscow, Russia\\
           email: artamon@sai.msu.ru}
             
\date{Received ...; accepted ...}
                            
\authorrunning{V.Vakulik et al.}
\titlerunning{Color effects in Q2237+0305}  

\abstract{Photometry of the Q2237+0305gravitational lens in VRI spectral  bands with the 
1.5-m telescope of the high-altitude Maidanak observatory in 1995-2000 is pre-
sented. Monitoring of Q2237+0305 in July-October 2000, made at  nearly daily basis, 
did not reveal rapid (night-to-night and intranight) variations of brightness of 
the components  during this time period. Rather slow  changes of magnitudes 
of the components were observed, such as 0.08m fading of B and C components 
and 0.05m brightening of D in R band during July 23 - October 7, 2000.
By good luck three nights of observation in 1999 were almost at the time 
of the strong brightness peak of image C, and approximately in the  middle of the 
ascending slope of the image A brightness peak. The C component was the most 
blue one in the system in 1998 and 1999, having changed its (V-I) color from 0.56m
 to 0.12m since August 1997, while its brightness increased almost 1.2m during this 
time period. The A component behaved similarly between August 1998 and 
August 2000, having become 0.47m brighter in R, and at the same time, 0.15m bluer. 
A correlation between the color variations and variations of magnitudes of the 
components is demonstrated to be significant and reaches 0.75, with a regression 
line slope of 0.33. A color (V-I) vrs color (V-R) plot shows the components settled 
in a cluster, stretched along a line with a slope of 1.31. Both slopes are noticeably 
smaller than those expected if a standard galactic interstellar reddening law were 
responsible for the differences between the colors of images and their variations 
over time. We attribute the brightness and color changes to microlensing of the 
quasar's structure, which we conclude is more compact at shorter wavelengths, as 
predicted by most quasar models featuring an energizing central source. 

}

\maketitle

\section{Introduction}
 
The Q2237+0305 gravitational lens (the Einstein Cross) is one of the most
impressive manifestations of the gravitational lensing phenomenon - four images
of the same high-redshift quasar ($z=1.695$) are arranged almost symmetrically
around the lensing galaxy nucleus ($z=0.039$) within a circle of approximately 
2\arcsec \enskip diameter. The Q2237+0305 system is an excellent target to 
study microlensing events, because the light beams, corresponding to the 4
lensed quasar's images, pass through the interior, heavily populated part of the lensing 
galaxy and thus, have a high probability to intersect a significant mass of
microlensing stars as they pass through the inner disc 
(Kayser \& Refsdal \cite{kay89}).

The system has been intensively examined since 1987, when the first published 
measurements of magnitudes of the individual lensed quasar components in $g, r$ 
and $i$ Gunn filters were made by Yee (\cite{yee88}). The first attempt to 
build the light curves of the four quasar components was made in 1991 by Corrigan 
et al. (\cite{cor91}). They brought together 
all the available Q2237 images of sufficient quality, taken with different 
telescopes and in a variety of passbands, - Mould $B, V$ and $R$ and Gunn 
$g, r$ and $i$, - and reprocessed with a single algorithm. Having used 
the multicolor photometry data for 33 normal stars, whose $(B-V)$ colors ranged 
from -0.3 to 1.5, they calculated the relevant color equations, which
allowed them to reduce all the observations to a single passband. Their $r$ 
Gunn and $B$ light curves cover the time period from September 1986 up to December 
1989, and include the first microlensing event observed in August 1988 by Irwin 
et al. (\cite{irw89}).

A further attempt to use all the available observational data for Q2237+0305   
was made in 1994 by constructing "differential" light curves, which were argued 
to be free from the effects of different spectral bands, technique of 
zero-pointing, and the quasar intrinsic brightness variations, (Houde et al. 
\cite{hou94}). In addition to the data of 1986-89, contained in (Corrigan et al. 
\cite{cor91}), other results were used, taken in 1990 and 1991 by Crane et al. 
(\cite{cra91}), Racine (\cite{rac92}), Rix et al. (\cite{rix92}), and Houde et 
al. (\cite{hou94}).

The first program of regular photometric monitoring of Q2237+0305 was started in 
1990 at the Nordic Optical Telescope, (\O stensen et al. \cite{ost96}). A large 
number of measurements of the four quasar components in $V, R$ and $I$ spectral 
bands during five years were obtained, which permitted construction, with the 
use of Corrigan et al. (\cite{cor91}) zero-pointing, of the historic light curves, 
covering 9 years of observations. 

No regular multi-filter monitoring of Q2237+0305 is reported between 1996 and 
1999, excepting our results of $VRI$ photometry for three nights on 17-19 
September 1995, (Vakulik et al. \cite{vak97}), and the similar results by 
Burud et al. (\cite{bur98}), obtained with the Nordic Optical Telescope for 
a close epoch, 10-11 October 1995. No night-to-night or intranight brightness 
variations of the four components have been found for these time periods, 
while a noticeable change in the component B color as compared to the 
observations by Yee in 1987 (\cite{yee88}) has been reported in both works. 
A short time-scale monitoring with the CFHT in June 14-16, 1992 should be 
mentioned here, (Cumming \& De Robertis, \cite{cum95}), which also did not 
reveal any photometric variations in $R$ and $I$ bands during a three-day 
period. 

In 1999 and 2000, the results of $VRI$ photometry in 1997 and 1998 with the 
Maidanak 1.5-m telescope were published, (Bliokh et al. \cite{bli99}) and 
(Dudinov et al. \cite{dud00}). Recently, the superb results of a detailed 
long-term monitoring, obtained within the OGLE program from 4 August 1997 to 
5 November 2000 has become publicly accessible, and partly presented in two 
papers of Wozniak et al. (\cite{wo100}) and (\cite{wo200}). And in 2002, 
the results of monitoring of Q2237+0305 by GLITP collaboration appeared, 
which cover the 4-month period  October 1999 - February 2000 (Alcalde et al. 
\cite{alc02}). The most recent publication of the results of low-resolution 
observations with the 3.5-m telescope at the Apache Point Observatory should 
be also mentioned here, (Schmidt et al. \cite{sch02}), which have given the 
Gunn $r$ lightcurves of A and B components for 73 dates between July 1995 
and January 1998. In spite of a rather low photometric accuracy, - an error 
bar of $0.1^m$ to $0.2^m$ is reported, - the data are of value first of all 
because they include the brightness peak of A component in 1996. As far as 
we know, no more data about this event have been ever published, though 
reported in private communication, (e.g. R. \O stensen). 

The brightness records taken in broad-band filters are a good starting point 
for theorists to estimate the size of the quasar radiating region in the
visual (ultraviolet rest frame) continuum, and to determine the range for 
microlens masses responsible for the observed brightness variations. Both 
the statistical analysis of long-term monitoring data, and simulation of 
the isolated microlensing peaks have been applied to calculate these values, 
(Nadeau et al. \cite{nad91}, Lewis \& Irwin \cite{lew96}, Refsdal \& Stabell 
\cite{ref93}, Webster et al. \cite{web91}, Wyithe et al. \cite{wwt01}, Wyithe 
et al, \cite{wwt02}, Wyithe et al. \cite{wyi02}, Yonehara \cite{yon00}), 
having given the estimates of the quasar dimension, -  $10^{15}cm$ to 
$10^{16}cm$ in the optical continuum, and a great variety of microlens masses 
ranging from $0.0006M_\odot < M < 0.006M_\odot,$ (Nadeau et al. \cite{nad91}), 
to $0.1M_\odot$, (Wambsganss \cite{wam92}), all indicating however, that 
microlensing events in the system are mostly caused by subsolar-mass objects. 

These available models confront a problem which probably indicates a
breakdown of the simple assumption that the luminous quasar is a uniformly
bright accretion disc. The dilemma is that the short time scale of the
observed events like the C image peak in July 1999 and the A brightness
peak in November 1999 occur on such short time scales that small accretion 
disc diameters are implied; such small bright accretion discs would have 
occasional strong brightness peaks of several magnitudes that are never observed. 
This probably tells us that 
a more complex model where the time scale of the brightness peaks is related 
to a quasar structure crossing time, not the crossing time of the entire 
quasar luminosity. We expect to apply, in a subsequent report, the Schild and
Vakulik (\cite{sv03}) double ring quasar model that successfully models the 
long history of Q0957+561 microlensing observations.

No systematic multicolor photometric measurements existed until the monitoring
program with the NOT was started in 1990 (\O stensen et al. \cite{ost96}). 
Meanwhile, a suspicion was expressed by Corrigan et al. in 1991, and in 1992 
by Rix et al. independently (Corrigan et al. \cite{cor91}, Rix et al. 
\cite{rix92}), that the color indices of the components might have changed 
since the first three-color observations by Yee, (\cite {yee88}). It was 
a very important statement, since reducing different datasets to a single 
light curve, (e.g. Houde et al., \cite{hou94}), 
as well as determining the extinction law in the Q2237+0305 lensing galaxy, 
(Yee \cite{yee88}, Nadeau et al. \cite{nad91}, Falco et al. \cite{fal99}),
are substantially based on the assumptions, that 
"all four components have identical intrinsic color indices", and "their 
observed color differences are due to different degrees of interstellar 
extinction and reddening by the same extinction law", (Houde et al. \cite{hou94}),
and "the magnification is wavelength independent... and time independent", 
(Falco et al. \cite{fal99}). In particular, Falco et al. (\cite{fal99}) 
measured the value of $R_V$ for Q2237+0305 to be equal 5.3 and came to a 
conclusion about great differences in the extinction laws for lensing 
galaxies from a sample consisting of 23 gravitational lens systems. 
However, as it can be seen from Fig. 4 in Falco et al. (\cite{fal99}), 
the differences may be significant at wavelengths shorter than 550nm, while 
at larger wavelengths the difference between the extinction curves does not exceed
the error bars.

 In discussing the results of the five-year $VRI$ monitoring of Q2237+0305, 
 \O stensen et al. (\cite{ost96}) did not analyze, however, any color changes 
 in the system, having noted only "very nearly equal" colors for the components 
 A and B, as well  as roughly equal colors of C and D, with the extinction 
 difference between the  pairs of $0.6^m$ in $V$ band, provided the extinction 
 law follows $\lambda^{-1}$,  according to Houde et al. (\cite{hou94}). 
 Meanwhile, Vakulik et al. (\cite{vak97}) and Burud et al. (\cite{bur98}) 
 reported that the B component became the most blue one in the system in 1995, 
 as compared to observations by Yee in 1987, (Yee \cite{yee88}).

The next step in determining colors and color changes in the Q2237+0305 system 
was made in $VRI$ observations with the Maidanak 1.5-m telescope in 1997-1998, 
presented in Dudinov et al. \cite{dud00} and Dudinov et al. \cite{dud01}. 
Variations of colors were argued to be significant, and a tendency of the 
components to become bluer as their brightness increased has been demonstrated 
with the use of all available multicolor data. Unfortunately, the remarkable 
monitoring by Wozniak et al. \cite{wo100} and Wozniak et al. \cite {wo200} has 
been made only in $V$ band, and thus can not be used to investigate the
color changes, while the most recent data of the GLITP collaboration have been 
taken in $V$ and $R$ filters for a campaign of 4 months only (Alcalde et al. 
\cite{alc02}).

By this time, a great amount of observations of Q2237+0305 in spectral 
ranges other than visual continuum exists, - VLA observations at 20cm and 
3.6cm (Falco et al. 1996), observations in the near and mid-IR (Nadeau 
et al. \cite{nad91}; Agol et al. \cite{ago00}), and in the quasar emission 
lines (Fitte \& Adam \cite{fit94}; Racine \cite{rac92}; De Robertis \& Yee 
\cite{rob88}, Lewis et al. \cite{lew98}, Saust \cite{sau94}). The observed 
magnitudes of the components have been found to be almost unaffected by 
microlensings in these spectral ranges, which indicates that much larger 
quasar features radiate in IR and in the radio, as well as in the emission 
lines, as compared to the optical continuum. The recent detection of an 
arc of C III] emission, connecting A, D and B components (Mediavilla et al. 
\cite{med02}), should be regarded as a visual proof of the extended emission 
line region of the source. Because of the low sensitivity of a large source 
brightness to microlensing, the brightness ratios for the components, 
measured in these spectral ranges, were used to test the validity of a 
great variety of the existing macrolensing models, listed by Wyithe et al. 
(\cite{wyi02}).

Observations in UV with the HST (Blanton et al. \cite{bla98}) and the recent 
X-ray imaging of Q2237+0305 with the Chandra X-ray Observatory 
(Dai et al. \cite{dai03}) should be also mentioned here, which provided, 
in particular, highly accurate relative coordinates of the components 
(Blanton et al. \cite{bla98}) and   the upper limits for the physical size 
and brightness of the Broad-Line Region producing Ly-$\alpha$ emission,  
(Dai et al. \cite{dai03}). Also, the Chandra data permitted calculation of the 
time delay between the A and B components of 2.7 hours.

\section{Observations}

Our observations were carried out with the 1.5-m AZT-22 telescope of the 
high-altitude Maidanak observatory, (Central Asia, Republic of Uzbekistan), 
known for its superb seeing conditions and a large number of cloudless nights,
(Ehgamberdiev et al. \cite{ehg00}). Because of technical reasons, we had 
to use three different CCD cameras in our observations, – Pictor-416 camera 
in 1995, Pictor-416 and TI 800 x 800 cameras in 1997 and 1998, and ST-7 
camera in 1999 and 2000.  And because of technical reasons again, both 
f/8 and f/16 focal lengths were used in observations. The LN-cooled TI 
800 x 800 camera, with pixel size of $15\mu$, kindly provided by 
Prof.D.Turnshek, unfortunately revealed some peculiarities, caused by 
the charge transfer inefficiency, that is characteristic for the CCD's 
of this generation, (Turnshek et al. \cite{tur97}). In particular, 
noticeable stretching of stellar images in the direction of charge 
transfer is observed, as well as a dependence of the PSF upon coordinates 
at the chip plane. In addition, sensitivity irregularities of the chip 
can not be corrected  satisfactorily, with the output of the flatfielding 
procedure dependent on the signal level. All these peculiarities reduced 
the actual accuracy of photometry, that is seen in Tables ~\ref{R95-99} 
and ~\ref{vri95-00}. 

Unfortunately, a poor telescope tracking system 
spoiled the intrinsically good seeing of Maidanak site sometimes and
did not permit use of exposures 
longer than 3 minutes. To provide sufficiently high accuracy of our photometry 
with such short exposures, we took images in series, consisting of 10 to 
20 frames each. The frames were averaged before being subjected to photometric 
processing, while a comparison of photometry of individual frames enabled us 
to obtain an adequate estimate of the random error inherent in a particular series.

Most of images has been taken in $R$ band, - 31 dates in 1995-1999, 
(Table~\ref{R95-99}), plus 46 dates in 2000, (Table~\ref{R2000}) - which were 
obtained almost at a daily basis during 2.5 months. There is also photometry 
in $V$ and $I$ bands for 17 dates in 1995-2000, (Table~\ref{vri95-00}). Some 
results have been presented in our previous publications (Vakulik et al. 
\cite{vak97}, Bliokh et al. \cite{bli99}, Dudinov et al. \cite{dud00}, 
Dudinov et al. \cite{dud01}). We present here the results of all our 
observations, including those which have been never published.
In particular, the observations of July-October 2000 are presented, which
have been undertaken to search for short-period (night-to-night) 
variations of brightness. The appearance of the Einstein Cross at six epochs
between October 1995 and August 2001 can be seen in Fig. 1, which 
clearly demonstrates high photometric variability in the system.

In addition to magnitudes of the components, the seeing conditions are also 
presented in Tables~\ref{R95-99} and ~\ref{R2000}, - the values of FWHM for 
particular nights, the scales and the CCD camera used.

\section{Photometric Reductions}

The difficulties inherent in accurate photometry of ground-based 
images of Q2237+0305, have been noted by many authors, (Burud et al. 
\cite{bur98}, Corrigan et al. \cite{cor91}, Vakulik et al. \cite{vak97}, 
Yee \cite{yee88}). They are due mainly to its extremely compact spatial 
structure, with the wings of the quasar images overlapping even under good 
seeing conditions. Additional difficulties are due to the presence 
of a rather bright foreground lensing galaxy, with its steep radial 
brightness distribution. These are the main reasons for poor agreement 
of the results of different monitoring programs, and even for a noticeable 
discrepancy in photometric results for the same data reduced with  
different algorithms, (Burud et al. \cite{bur98}, Alcalde et al. \cite{alc02}).
In photometry of the data of 1995-1999, we used the method described in 
Vakulik et al. \cite{vak97}, and Bliokh et al. \cite{bli99}, 
which is in general 
features similar to the double iterative PSF subtraction method, proposed 
by Yee (\cite{yee88}), who was the first to present spatially resolved 
photometry of the system. In short, the method consists of the following.

The PSF estimate is obtained from a reference star image, and is further 
superimposed upon each image component and the galaxy nucleus alternately, 
and then subtracted in such a way, that no depressions would appear in the 
residual brightness distribution. Such a procedure is repeated iteratively 
until a stable convergence of estimates of brightness and coordinates of the 
components is achieved. Then, according to the resulting estimates, the 
quasar components are subtracted, and the residual galaxy brightness 
distribution is smoothed with a rather broad median filter. After the 
resulting galaxy brightness distribution is subtracted from the initial image, 
removal of quasar components is repeated, followed again by smoothing the 
galaxy brightness distribution with a successively decreasing window. The 
iterative process stops when the width of the median filter becomes of 
order of the PSF width.
 
In processing our data for the 2000 observing season, another method was 
applied, which used the known relative coordinates of the components and 
an analytical model of the brightness distribution in the galaxy, represented 
as a sum of three two-dimensional Gaussian functions. Before describing the 
algorithm, consider the basic principles of photometry for compact groups 
of star-like objects, that have been implemented in the known algorithms
of other authors.

Even in the images of Q2237+0305 taken with the Hubble Space Telescope, the 
quasar components are star-like and thus, in the isoplanacy 
region, the entire picture (photometric model of the system) can be represented 
as a sum of the PSF's $r(x-x_k,y-y_k)$, and the galaxy light distribution $g(x,y),$
and in the case of a sampled CCD image, may be written as:
\begin{equation}
\label{ph_model}
f(i,j)=\sum^4_{k=1}I_k r(i-x_k,j-y_k)+g(i,j),
\end{equation}
where $i$ and $j$ are pixel numbers in $x$ and $y$ axes, chosen in parallel 
 to the CCD lines and columns, respectively. The unknown parameters, - the 
 coordinates of the components in the detector reference frame, $x_k, y_k$, 
 their relative brightnesses $I_k$, and the galaxy light distribution $g(i,j)$, -
 are usually estimated from a requirement to minimize the difference between a 
 model and the observed brightness distribution in the detected 
 image according to some 
 criterion, -  e.g. the minimum of the sum of square residuals criterion: 
\begin{equation}
\label{func}
\Phi (\vec p)=\sum_i \sum_j(F(i,j)-f(i,j, \vec p))^2 = min.
\end{equation}  
Here $F(i,j)$ is the brightness distribution in the detected image, and the 
set of unknown parameters is denoted as $\vec p$ for short. The estimate of 
the PSF can be obtained from the images of reference stars near the object.

As was noted above, noticeable difficulties in photometry of Q2237+0305 
components are caused by the foreground lensing galaxy, with its light 
distribution $g(x,y)$ being unknown. In minimizing Eq. \ref{func}, or another 
one similar to it, - the galaxy brightness distribution is usually represented 
either analytically, - (e.g. Burud et al. \cite{bur98}, Alcalde et al. 
\cite{alc02}), or its digital form $g(i,j)$ is estimated, - e.g. the MCS 
algorithm (Magain et al. \cite{mag98}, Burud et al. \cite{bur98}). 

To solve the problem, iterative algorithms are often used, which in fact 
approximately realize minimization of Eq. \ref{func}, and also permit to obtain the 
estimate of $g(x,y)$ either analytically (Teuber \cite{teu93}, Ostensen et al.
\cite{ost96}) or in a digital form, (Yee \cite{yee88}, Vakulik et al. \cite
{vak97}). The resulting analytic or numerical model can be treated further 
in photometry of Q2237+0305 components as a known function. Such an approach 
noticeably simplifies  the solution procedure, and provides good intrinsic 
convergence, (Corrigan et al. \cite{cor91}, Alcalde et al. \cite{alc02}, 
Burud et al. \cite{bur98}), but unfortunately, does not ensure the absence 
of systematic errors in estimating the magnitudes of the components caused 
by an inadequate galaxy model. 
 
A new image subtraction method proposed by Alard \& Lupton (\cite{ala98}) 
and successfully applied by Wozniak et al. (\cite {wo100}, \cite {wo200}) and 
by Alcalde et al. (\cite{alc02}) in Q2237+0305 photometry, is seemingly free 
from this weak point. However, a comparison of photometry results for 
Q2237+0305 published by the OGLE group and those obtained with other methods, 
reveals some systematics in the components magnitudes, that is probably caused 
by a bias of brightness estimates in their reference image.

The PSF is usually represented either numerically, or as an analytic function. 
In this work, the following approach was used. We transformed all the detected 
images to the same axisymmetrical Gaussian PSF with a preassigned parameter 
$\sigma_s$ using the inverse linear filtration procedure: 
\begin{equation}
\label{fourier}
F(i,j)=\tilde W \left({{\tilde F_0(\omega_l, \omega_n)\cdot R(\omega_l,\omega_n)}
\over{\tilde r(\omega_l, \omega_n)}}\right),
\end{equation}
Here $\tilde F_0(\omega_l,\omega_n)$ is the Fourier transform of the initial image, 
$F(i,j)$ is the transformed (standardized) image, and $\tilde W$ is the inverse 
Fourier transform operator. A complex-valued inverse filter $w(\omega_l, \omega_n)
=1/{\tilde r(\omega_l, \omega_n)}$ is composed from the Fourier transform of the 
initial PSF $r(i,j).$ A function $R(\omega_l,\omega_n)=exp[-\sigma_s^2(\omega^2_l
+\omega^2_n)/2]$ forms the Fourier spectrum of the standardized image with the 
Gaussian PSF for the given parameter $\sigma_s.$ To construct the inverse filter, 
a reference star about 64\arcsec south-west from the quasar, denoted as $\alpha$ 
star in Corrigan et al. (\cite{cor91}) was used.

In doing so, we did not try to noticeably increase the resolution in the initial 
images, and used a  transformation (\ref{fourier}) that is a linear one, and, in
contrast to non-linear filtration methods, retains photometric accuracy. To 
exclude dependence of the resulting PSF on the signal-to-noise ratio in the Fourier 
spectrum of a specific image, we also did not use any optimizing algorithms of 
image reconstruction, such as e.g. the well-known Wiener filtering.

Since the restoring filter is normalized to unity at the zero spatial frequency, 
such a transformation retains the integral brightness of an image, and thus the 
estimates of the components' brightnesses can be made in the units of the reference 
star brightness.

With such standardized images created, the sum in Eq.\ref{ph_model} can be 
represented as
\begin{equation}
\label{PSF}
s(i,j)=\sum^4_{k=1}I_k\exp \{-[(i-x_k)^2+(j-y_k)^2]/2\sigma_s^2\},
\end{equation}
where $\sigma_s$ is an effective width of the resulting PSF.

The distribution of light over the galaxy was represented by a sum of three 
two-dimensional Gaussian functions: 
\hspace{3cm}
$$g(i,j)=$$\hfill
\vspace{0.5cm} 
\newline
\vspace{-1.2cm}
$$=\sum^3_{m=1}I_m\exp\{-[(i-x_g)\cos\varphi_m 
+(j-y_g)\sin\varphi_m]^2/2\eta^2_m -$$ 
\vspace{-0.2cm}
\begin{eqnarray}
\label{gal_mod}
-[-(i-x_g)\sin\varphi_m+(j-y_g)\cos\varphi_m]^2/2\varepsilon^2_m\}, 
\end{eqnarray}
where $x_g,y_g$ are coordinates of the galaxy center, $I_m$ are normalizing 
coefficients, $\eta_m$ and $\varepsilon_m$ are parameters determining the 
characteristic widths of the Gaussian profiles along the major and minor axes 
respectively, with their meaning understood from Eq. 5, and finally, 
$\varphi_m$ defines the major axes orientation. 
Therefore, the photometric model of the system $f(i,j,\vec p)$ in Eq.\ref{func} 
can be represented as a sum of two constituents, $s(i,j)$ and $g(i,j)$, which 
describe the quasar components (Eq.\ref{PSF}), and a photometric model of the 
light distribution in the lensing galaxy, (Eq. 5).
A set of 26 unknown parameters denoted as $\vec p$, consists of four pairs of 
coordinates $x_k,y_k$ and coordinates of the galaxy center $x_g, y_g$, 
normalizing multipliers $I_k$, $I_m$, the parameters $\eta_m, \varepsilon_m,$ 
and orientations of axes $\varphi_m$ of the three Gaussian components of the 
galaxy model.

To calculate the parameters of the galaxy photometric model, as well as the 
coordinates of the quasar components, a set consisting of 14 best quality 
images was selected that was obtained on September 2, 2000 in $R$ filter under the 
atmospheric seeing of 0\farcs8 and better. The images were averaged and reduced, 
through the inverse linear filtration procedure described above, to the Gaussian 
PSF with $\sigma_s=0\farcs34$, (FWHM of 0\farcs8).

The least-squares algorithm was used to calculate the brightnesses and coordinates 
of the components and the parameters of the galaxy photometric model from the 
condition expressed by Eq. \ref{func}. It should be noted, that in such a way we 
obtain parameters of the galaxy model, that is the result of the convolution 
of an actual galaxy light distribution with the Gaussian PSF with the 
given $\sigma_s=0\farcs34$. Since Gaussian functions were adopted both for the PSF 
and for the constituents of the galaxy model, the deconvolved galaxy model parameters 
can be easily calculated. Such deconvolved parameters are presented in Table~\ref{gal_mod}.

In Table~\ref{rel_pos}, the relative positions of the B, C, D components and the 
galaxy center in the equatorial coordinate system, calculated from the 14 selected 
images with the procedure described above are presented. Our coordinates agree 
within 0\farcs015 with those obtained from the HST images (Crane et al. \cite{cra91}, 
Blanton et al. \cite{bla98}). 
\begin{table}
\caption{Parameters of the photometric model of the lensing galaxy for 
Q2237+0305 system.} 
\label{gal_mod}
\begin{center}
\begin{tabular}{ccccc}
\hline
 \noalign{\smallskip}
 $m$  & $I$ & $\eta(\arcsec)$& $\varepsilon(\arcsec)$& P.A.(\degr)\\
 \hline
 1  &$0.875\pm0.021$ &$0.264\pm0.045$  &$0.206\pm0.038$  &$57\pm3$\\
 2  &$0.090\pm0.011$ &$1.260\pm0.075$  &$0.790\pm0.041$  &$78\pm2$\\
 3  &$0.035\pm0.002$ &$5.440\pm0.530$  &$2.840\pm0.110$  &$58\pm4$\\
\hline
\end{tabular}
\end{center}
\end{table}
\begin{table}
\caption{Relative angular positions of Q2237+0305 A,B,C,D components and the 
galaxy center (G) from observations of 2000.} 
\label{rel_pos}
\begin{center}
\begin{tabular}{ccc}
\hline
\noalign{\smallskip} 
Component & $\Delta\alpha(\arcsec)$ & $\Delta\delta(\arcsec)$\\
\hline
A   &$0.000$              &$0.000$ \\
B   &$-0.674\pm0.003$     &$1.679\pm0.004$\\
C   &$\quad 0.624\pm0.005$&$1.206\pm0.004$\\
D   &$-0.867\pm0.008$     &$0.513\pm0.003$\\
G   &$-0.085\pm0.014$     &$0.939\pm0.006$\\
\hline
\end{tabular}
\end{center}
\end{table} 

In the subsequent photometric processing of all the available data, every 
image was reduced to a "standard" PSF, and the corresponding quasar image 
brightnesses were estimated by minimizing the function (\ref{func}), with 
the parameters of the galaxy model and the relative coordinates of the 
components being fixed, according to Tables ~\ref{gal_mod} and ~\ref{rel_pos}.
The $\alpha$ star from Corrigan et al.(\cite{cor91}) was used as a secondary 
photometric standard, with its magnitudes taken from this work. 

Photometry of the image sets taken during a single night does not show brightness 
variations that might be regarded as significant as compared to the 
photometry uncertainties. Therefore, the brightness estimates taken within 
a night were averaged, and the formally calculated error in the mean can be 
regarded as a measure of the inner convergence of our photometry. The method 
ensures photometry with no seeing-dependent systematic errors, inherent
in some other methods, - for images with a PSF up to 1\farcs4.

\section{Results of $VRI$ photometry}

Our photometry is presented in Tables~\ref{R95-99}, ~\ref{R2000}, and 
~\ref{vri95-00}. The magnitudes were zero-pointed with Yee's (\cite{yee88}) 
reference star, with its magnitudes taken from Corrigan et al. (\cite{cor91}).
Our measurements in $R$ band in 1997-2000 are plotted in Fig. 2, 
where the OGLE data (Wozniak et al \cite {wo200}) taken in $V$ filter are shown 
in grey. 
For better comparison, our data of Tables 4 and 5 are shifted by small amounts 
along the vertical axis, -  0.1, 0.13, 0.15 and 0.3 magnitudes for A, B,C and D, 
respectively. 

The most important brightness changes observed were:

\begin{enumerate}

\item An increase of the image A brightness, starting at the end of 1998 and 
peaking, according to photometry of Wozniak et al.(\cite{wo200}) and Alcalde 
et al. (\cite{alc02}), in the middle of November 1999. We observed almost $0.4^m$ 
brightening of A image between our observing seasons in 1998 and 1999.

\item A monotonic decline of almost $1.0^m$ in image B brightness starting with 
our earliest, 1995.8 observation. It has become the faintest component in $R$ 
band by September 2000.

\item A strong brightness peak in image C. Our observations in July 19-22, 1999 
were made near the brightness peak of the C component, seen in the 
well-sampled light curves of Wozniak et al. (\cite{wo200}). The C image became 
almost $1^m$ brighter in $R$ band between August 1997 and July 1999. Thus we 
have an excellent occasion to detect the color change that accompanied the 
brightness peak. 

\item A noticeable growth of the D image brightness, which is no more the faintest 
one since September 2000.

\end{enumerate}

Our measurements, presented in Fig. 2 are in a good qualitative agreement 
with more detailed and accurate single-filter light curves of Wozniak et al. 
\cite{wo200}, taken in $V$ band for a similar epoch. A large scatter of points 
for 1997 and 1998 in Fig. 2 are due to the dates, when the TI 800 x 800 
CCD camera was used. We compared 
our $V$ magnitudes, taken with the TI 800 x 800 camera in 1997-98, (Table~\ref{vri95-00}) 
with the same dates of OGLE monitoring, and found that the OGLE $V$ magnitudes are 
systematically smaller than our measurements with this camera. The greatest differences 
are for the A and D components, reaching approximately $0.2^m$, with $0.1^m$ for B and C.
 As seen from comparison of Table~\ref{R95-99} with Table~\ref{R2000}, 
where the photometry with the ST-7 camera is presented, the latter is almost 
an order of magnitude more accurate as compared to the TI 800 x 800 data. 
\begin{table}
\caption{A comparison of three programs of Q2237+0305 photometry: Maidanak (this work,
ST-7 CCD), OGLE and GLITP; $V$ magnitude differences for A,B,C and D components; 
observations of 2000.} 
\label{comp}
\begin{center}
\begin{tabular}{ccccc}
\hline
\noalign{\smallskip} 
Programs     & $\Delta V_A$ & $\Delta V_B$ & $\Delta V_C$ & $\Delta V_D$\\
\hline
GLITP - OGLE & 0.07 & 0.01  & 0.14  & -0.18\\
Maidanak-OGLE& 0.06 & 0.05  & 0.07  & -0.15\\
\hline
\end{tabular}
\end{center}
\end{table} 

Our well-sampled and most accurate measurements, made in July-September 2000 
with the ST-7 camera, - the datapoints near the right edge of Fig.~\ref{R97-2000}, 
- can be seen in Fig. 3 more in detail, (see also Table~\ref{R2000}). 
For better clarity, the light curves in Fig. 3 were arbitrarily shifted 
along the magnitude axis, and fit with quadratic polynomials, with the 1\% 
error strips shown. Variations of brightness of all the components were moderate 
during this time period,  – about $0.02^m \div 0.03^m$ per month, – and may be 
approximated by the second-order polynomials quite well. The brightness 
estimates for the A component are mainly within a 1\% deviation with respect to 
the fitted curve. A correlation between the rapid brightness variations of all 
the components seen in Fig. 3 could be ascribed to quasar intrinsic 
brightness changes, except that since their amplitudes are larger for the 
fainter components, these variations are probably not real and are more
likely due to errors.

We compared our photometry of July-September 2000 in $V$ band with the OGLE 
data, obtained for the same dates, and, since our data do not overlap with 
the observations of GLITP collaboration, we made a similar comparison between 
their photometry and that of OGLE. The results of such a comparison are presented 
in Table~\ref{comp}. Here, positive differences mean that OGLE magnitudes 
are smaller. The difference between our photometry and that of OGLE program 
will be even smaller for A, B and C images if one takes into account $0.034^m$ 
difference in magnitude for $\alpha$ star adopted in Wozniak et al. (\cite{wo100}) 
and in this work, though systematics for the C component will become larger. 

\section{Variations of color in Q2237+0305}

The first multicolor observations by Yee (\cite{yee88}) have immediately shown, 
that the components differ in their colors.
An obvious dependence of the components' reddening on the distance to the 
galaxy nucleus allowed Yee to explain it by selective extinction in the dusty 
matter of the lensing galaxy. This suggestion made it possible to estimate 
the extinction law in the lensing galaxy, which, according to Nadeau et al. 
(\cite{nad91}) and Yee (\cite{yee88}), is similar to that in our Galaxy. 
It should be emphasized here, that the conclusion was based on the analysis 
of color differences of the components for a fixed epoch. 

As mentioned in the Introduction, a suspicion arose in 1991 and 1992, that 
the colors of the components might have changed, (Corrigan et al. \cite{cor91}; 
Rix et al. \cite{rix92}). In particular, Corrigan et al. did not find any 
significant variations of $(B-r)$ colors of the components with time, but they 
were the first to notice that "there may be a small color change in image A as 
the $r$ magnitude gets fainter" (Corrigan et al. \cite{cor91}). They referred to 
the work by Wambsganss \& Paczinski (\cite{wp91}), where the possibility is 
discussed that, if the quasar structure is wavelength dependent, microlensing 
events will differently reveal themselves in different spectral regions. 
In particular, according to Wambsganss \& Paczinski (\cite{wp91}), the bluer 
inner parts of the continuum source might be more strongly amplified as 
compared to the outer parts. 

Rix et al. (\cite{rix92}), analyzing their observations in $U$ and $R$ bands 
with the Hubble Space Telescope, plotted their $(U-R)$ colors against $(g-i)$ 
colors of the components, measured by Yee (\cite{yee88}), and concluded that 
they "are only marginally consistent" with the reddening line derived by Nadeau 
et al. (\cite{nad91}). They suggested, the discrepancy could be due to either 
variable dust extinction in the lensing galaxy, or to the effects of microlensing 
color changes, first noted by Kayser et al. (\cite{kay86}) and later investigated
by Wambsganss \& Paczynski (\cite{wp91}) and Wambsganss (\cite{wam92}) in simulations. 

 We have already analyzed the behavior of the relative colors of the Q2237+0305 
 components qualitatively, (Dudinov et al. \cite{dud00} and Dudinov et al. 
 \cite{dud01}), based upon our observations on Maidanak in 1995 (Vakulik et al. 
 \cite{vak97}), and in 1997-1998, and also upon all available multicolor 
 observations by other authors, (Burud et al. \cite{bur98}; \O stensen et al. 
 \cite{ost96}; Rix et al. \cite{rix92}; Yee \cite{yee88}).  A tendency for the 
 components to become more blue as their brightness increases has been noted 
 there, but no quantitative relationships have been derived.

We present here our measurements of the colors of the A,B,C,D components, 
and the attempt to quantitatively analyze the behavior of $(V-R)$ and $(V-I)$ 
color indices of the components using our data taken in 1995-2000. $VRI$ 
photometry is presented in Table~\ref{vri95-00}, and $(V-R)$ and $(V-I)$ colors 
can be seen in Table~\ref{colors}. Formal errors for these quantities, calculated 
as the errors of the average, range from $0.02^m-0.03^m$ (A component) to 
$0.03^m-0.05^m$ (D component), – for the most accurate observations of 2000, – 
and, as seen from Table~\ref{vri95-00}, are within $0.08^m - 0.15^m$ for the 
observations of 1997-1998, made with the TI 800 x 800 camera. 

As shown in the previous section, our photometry is in quite satisfactory 
agreement with that obtained by other observers for close epochs, (e.g. 
Alcalde et al. \cite{alc02}, Woznyak et al. \cite{wo200}). At any rate, 
the discrepancy does not exceed that obtained when different algorithms are 
applied to the same data, (Alcalde et al. \cite{alc02}, Burud et al. \cite{bur98}).

However, one should keep in mind the peculiarities of Q2237+0305  photometry 
mentioned above in analyzing and interpreting the lightcurves in general, 
and especially those combined from heterogeneous observing data. With this in 
view, more weight should be given to the analysis of relative quantities, 
which are less sensitive to differences in observational circumstances and 
algorithms of image processing. In particular, relative colors and relative 
magnitudes, as well as their variations are such quantities. Examining their 
behavior in time, and their relationships with each other in microlensings 
can be a valuable source of additional information about the physical properties 
of both the quasar and lensing galaxy. In particular, they can be used to probe 
the spatial structure of the quasar at different wavelengths, (Wambsganss \& 
Paczynski \cite{wp91}), and to determine the extinction law in the lensing galaxy.

A correlation between $(V-I)$ colors  of the components and their $R$ magnitudes 
can be seen from Fig. 4, where the components are marked 
with different symbols. It is interesting to note, that B, C and D components 
are arranged just along a line in this diagram, while the A component forms 
a separate cluster of points. We can not refute the possibility of some systematic 
errors in our photometry, but we argue that they would hardly arrange the B, C 
and D components along a single line so well, - a correlation reaches 0.8 for them, - 
and separate the A component so significantly. Moreover, we studied the systematic 
errors of our algorithms very carefully in simulation and found, that their effect, 
if present, might only slightly bias the color of the C and D components to 
larger values, i.e. make them redder, as compared to A and B, (see Vakulik et al. 
\cite{vak97} for more details). We  see from Fig. 4, however, that 
the cluster of point for A image is shifted towards redder colors with respect 
to the cluster for B, C and D. 

If all the components were equally macroamplified, and if both the colour 
differences of the components and their variations in time were caused by 
 the interstellar reddening law, similar to that for our Galaxy, a linear 
 relationship between $(V-I)$ and $R$ could be expected, with a regression 
 slope of about 0.42 for $(V-I)$ base, (Schild~\cite{sch77}). Microlensing 
 events, with their still unknown brightness-color dependence, would disturb 
 and rearrange this order, making the components follow the reddening line 
in the average, but forming individual clusters of points, with the patterns 
and stretches, determined by the level of microlensing activity at a particular 
time period, and by the unknown character of color-brightness dependence of 
microlensings. 

However, the existing macrolens models predict different macroamplifications for
the components. According to the macromodel by Schmidt et al. (\cite{sch98}), rather 
well confirmed by the observations in emission lines (Fitte \& Adam \cite{fit94}; 
Racine \cite{rac92}; Lewis et al. \cite{lew98}, Saust \cite{sau94}), and in the IR 
spectral range, (Nadeau, et al. \cite{nad91}; Agol, et al. \cite{ago00}), where no 
microlensing effects are expected, the A, B and D components must be almost 
equally macroamplified, with the flux ratios of 0.25, 0.27 and 0.32 respectively. 
The C component is expected to have the least macroamplification factor by this model, 
- flux ratio of 0.15 is predicted for it. It means, that in this case the components 
can not be expected to sit along a line in the $(V-I)$ vrs $R$ plot. In the presence 
of microlensings, the components would produce a family of clusters, shifted with 
respect to each other by the amount of flux ratio differences.

We see in  Fig. 4 another situation however. While B, C and D 
components produce three overlapping clusters, all of them being stretched 
approximately in the same direction, the A datapoints form a separate cluster, 
stretched along a line with a slope similar to that of the joint B, C and D 
cluster. All the points in the joint cluster are rather well correlated, with 
a correlation index of $0.8 \pm 0.1$ and a regression line slope of $0.33 
\pm 0.08$. The points formed by the A component are also rather well correlated,
with a correlation index of 0.84 and a regression line slope of 0.36. 

To eliminate possible additive constituents of the color-magnitude dependence, 
which may differ for different components, and to focus on the analysis of changes, 
we studied a correlation between the deviations of $(V-I)$ colors from their 
average over the whole time period, and similarly calculated variations of brightness 
in $R$ band. The diagram can be seen in Fig. 5. The quantities are rather 
well correlated, with a correlation index of $0.75\pm0.08$, and a regression slope 
of $0.31\pm0.08$, in a good agreement with that of Fig. 4 for B, C and D 
components, but the A component is not situated separately this time. The uncertainties 
are given for an 80\% confidence interval. 

A diagram of color $(V-I)$ - color $(V-R)$, which is known to be of great diagnostic 
importance for the study of dust extinction, is usually presented by all 
authors of multicolor observations, e.g., Yee (\cite{yee88}), Rix et al. (\cite{rix92}), 
Burud et al. (\cite{bur98}), but, as was noted above, only for a fixed epoch. 
In such a diagram, the color indices should be proportional to each other for any 
color base and for any type of dust extinction, with the slope determined by the 
reddening law. The diagram, built with the use of our measurements (see Table
~\ref{vri95-00}), can be seen in Fig. 6. A large range of color variations  
of the C component should be particularly noted. It is quite real and can be 
explained by the presence of observations of 1999 in our data, - two asterisks 
near the origin. As was noted, these data were obtained near the July 1999 brightness 
peak of C, when it became almost $1^m$ brighter during two years, (Wozniak et al. 
\cite{wo200}), and exceeded the B component in brightness. The datapoints in this 
diagram are found along a line with a slope of approximately $1.31\pm 0.14$, which 
is much less than 2.15 for these color indices expected for the interstellar 
reddening law in our Galaxy, (Schild \cite{sch77}). 
We conclude that if the extinction law in the lens galaxy is similar to that of our
Galaxy, the observed color changes can not be explained by variable interstellar 
reddening.

\section{Color Changes Associated with Brightness Peaks}

A further perspective of the nature of the observed color changes comes from 
a comparison of the history of brightness changes with the history of color 
changes. The general features of long-term variations of the components 
magnitudes during 1995-2000 in comparison with the simultaneously determined 
colors can be seen in Fig. 7. Since durations of our observing  
seasons are small as compared to the characteristic time scale of the long-term 
variations of colors of the components, we calculated the mean values for 
$(V-I)$ and $R$ for every season, and plotted them as a function of time, 
(the midpoint dates of each season are used here).

Long-term variations of colors of the components are clearly seen in this figure, as 
well as a tendency of the components to change their color indices towards smaller 
values (bluer color) as their brightness increases. But the tendency is not always straightforward.
Some reddening of the components, preceding the subsequent decrease of their color 
indices at the stage of component brightening can be also seen. As noted above,
our observations in July 1999 were made very close to the brightness peak of C component, 
that is seen very well from the more complete and well sampled light curves of 
the OGLE program, Wozniak et al. (\cite {wo200}), while the A component was just 
in the middle of the ascending slope of its peak at this time, according to the 
observations of OGLE and GLITP programs, (Wozniak et al. \cite {wo200}, Alcalde et al. 
\cite{alc02}).

The relationship between brightness change and color is obvious and
approximately as expected from models of Wambsganss \& Paczynski (\cite{wp91}).
The most direct correlation is found for image C, where $(V-I)$ color is almost
perfectly anti-correlated with brightness. Thus as the C quasar image
brightened by almost $1.0^m$ in $R$ between 1997.5 and 1999.6, it became 
bluer by 0.42 magnitudes in $(V-I)$ color index (Fig. 7). A second interesting 
behavior is seen in the brightness of image A, where
we find that the brightness increased by $0.45^m$ as the color became 
bluer by $0.15^m$ in the $(V-I)$ color index. 
Just as interesting is the color history for image B, which underwent
a sustained slow brightness drop of $1.0^m$ during our monitoring
period. As it gently declined in brightness, it became $0.25^m$ redder 
in $(V-I)$ color from 1995.7 to 1998.8, and then again became 10\% 
($0.1^m$) bluer as the brightness continued to fade from 1998.8 to 2000.9.

We have already shown that this is not likely to be produced by a hole 
appearing in some absorbing clouds. If instead we view the image C brightness 
peak as a microlensing artefact where a compact object (star) in the lens 
galaxy passed in front of the quasar and caused the temporary brightening, 
we can compare to the calculations in Figs 1, 2, and 3 of Wambsganss \& 
Paczinski (\cite{wp91}). Their models were crafted to apply to Q2237, and 
they show approximately the correct brightness change ($1^m$ increase in $V$) 
and color change ($0.4^m$ bluer in $(B-R)$) for events with 1 year duration, 
and appear similar to the Wambsganss \& Paczinski (\cite{wp91}) Fig. 2 i,j 
pattern of a quasar image passing outside a cusp of a microlensing star.
We do not press these calculations further because we feel that the failed 
Wyithe, Turner, and Webster (\cite{wtw00}) prediction of a subsequent large 
brightness change invalidates all models with such simple accretion disc 
approximations. However almost any quasar model with an energizing central
source produces quasar structure which is more compact at shorter
wavelengths.  We expect 
to produce separately a series of models that can reproduce the observed
effects, based upon the double-ring Schild \& Vakulik (\cite{sv03}) 
model. 

Although quasar emission lines contaminate the color photometry in the
continuum-dominated filter bands, we doubt that the emission lines are
responsible for the large brightness-color effects found here, given that 
the large brightness changes observed are always associated with
microlensing of the quasar continuum.

\section{Conclusions}

\begin{enumerate}

\item Our observations demonstrate drastic changes of the 
component magnitudes, 
which are inherently uncorrelated in this system, confirming high probability
for microlensings, predicted for Q2237+0305 in 1989, (Kayser \& Refsdal 
\cite{kay89}). The highest gradient of brightness change was observed for 
the C component between 1997 and 1998, - almost $0.07^m$ per month in our $R$ 
filter. Almost the same value has been measured by OGLE program in their 
$V$ band for the same time period. However, a much more rapid 
brightness change 
of the C component was detected immediately after its extraordinary brightness 
peak in July 1999 by the OGLE program, - almost a $0.2^m$ decrease per month 
in their $V$ band, (Wozniak et al. \cite{wo200}).

\item No strong microlensing event occurred in the system during our detailed 
2.5-months monitoring in July-October 2000 but the fact that the B component 
has become the faintest one, after its long continuous fading beginning in
1995, (see Fig. 2 and Table~\ref{R2000}). No noticeable night-to-night 
brightness variations were detected in this time period. Moderate brightness 
changes were inherent in all the components, reaching a $0.03^m$ decrease per 
month for B and C, and $0.02^m$ brightening for D, (see Fig. 3 and 
Table~\ref{R2000}).

\item All the components demonstrated variations of their colors during 
1995-2000, which we argue to be real and significant. The most prominent 
change of color was observed for the C component, - $0.43^m$ for its $(V-I)$ 
color index during two years. The $(V-I)$ color indices of A, B and D 
were less 
variable during the whole time period, having changed from $0.3^m$ to 
$0.5^m$ for A, from $0.2^m$ to $0.5^m$ for B, and from $0.7^m$ to $0.45^m$ 
for D component, which became $0.2^m$ bluer between 1997 and 2000, having 
approached the B component in color and exceeded it in brightness.

\item The $(V-I)$, $(V-R)$ color-color plot shown in Fig.~\ref{vrvi} 
incorporates all our observations. The regression slope is $1.31\pm0.14$ for 
this diagram, i.e. much smaller than a value of 2.16, expected for the 
reddening line in our Galaxy for these color indices. We conclude that the
brightness and color changes observed are not caused by time variations in
reddening, but are more probably caused by microlensing of source structure
that is more compact at shorter wavelengths.

\item The large brightness peak of the C component in July 1999 was 
accompanied by large color change in the sense that as the C image 
brightness increased by almost $1^m$ both in our $R$ and Wozniak et al. 
(\cite{wo200}) $V$ bands, the color became bluer by $0.43^m$ in $(V-I)$. 
This is the sense and amplitude expected for microlensing of an object 
that is smaller at shorter wavelengths, and modelled previously by 
Wambsganss and Paczinski (\cite{wp91}). The colors of the other components 
behave similarly, though the amplitudes of their color variations are 
smaller, (see Fig. 7).  

\item Returning to Fig. 4, where the relationship between the
$(V-I)$ colors and $R$ magnitudes of the components is shown, we note
that the plot is inconsistent with the adopted models of macrolensing,
e.g. Schmidt et al. (\cite{sch98}). We think that most probably the A 
component is macroamplified almost $0.8^m$ more than B, C and D, which 
have almost equal amplifications.  

\end{enumerate}

We hope, that the data presented here will demonstrate the importance of   
multiband observations of gravitationally lensed quasars in general, and 
Q2237+0305 in particular. More detailed analysis of the obtained data 
and simulation with the new quasar structure model, will be presented in 
the next paper, which is in progress.

\begin{acknowledgements}

The authors thank the Maidanak Foundation, and its President Dr.Henrik N.
Omma personally for delivering the ST-7 CCD camera. We also appreciate a 
valuable financial support and kind attention to our work from \fbox{Dr.James Bush} 
and Prof. Kim Morla (Pontificia Universidad Catolica del Peru, Lima). 
The work has been also substantially supported by the joint Ukrainian-Uzbek
Program "Development of observational base for optical astronomy on Maidanak 
Mountain". The observations of 1997-98 have become possible thanks to funding 
from the CRDF grant UP2-302, with Prof. B.Paczynski as a US Co-Investigator, 
whom the authors from Ukraine greatly appreciate. The co-authors from 
Russia are also thankful to the Russian Foundation of Fundamental Research, 
grants  No.98-02-17490 and 1.2.5.5.
 
\end{acknowledgements}

\begin{table*}
\caption{Photometry of Q2237+0305 in $R$ band from observations with the Maidanak
1.5-m telesope in 1995-1999.} 
\label{R95-99}
\begin{center}
\begin{tabular}{cccccccc}
\hline
 \noalign{\smallskip}
Date & A & B & C & D & FWHM(\arcsec) & Camera & Scale(\arcsec /pix) \\ 
\hline
\hline 
95.09.17 &$17.18\pm0.03$ &$17.32\pm0.03$ &$18.13\pm0.06$ &$18.44\pm0.07$&0.90&Pictor&0.159\\           
97.07.02 &$17.10\pm0.05$ &$17.77\pm0.05$ &$18.10\pm0.07$ &$18.48\pm0.10$&0.73&TI&0.130\\
97.07.03 &$17.08\pm0.06$ &$17.62\pm0.06$ &$17.98\pm0.08$ &$18.55\pm0.12$&0.84&TI&0.130\\
97.08.29 &$17.15\pm0.03$ &$17.72\pm0.03$ &$18.08\pm0.12$ &$18.38\pm0.10$&0.85&TI&0.130\\
97.08.30 &$17.14\pm0.02$ &$17.75\pm0.03$ &$18.08\pm0.04$ &$18.41\pm0.05$&0.71&TI&0.130\\
97.08.31 &$17.16\pm0.02$ &$17.78\pm0.03$ &$18.05\pm0.05$ &$18.49\pm0.06$&0.81&TI&0.130\\
97.09.01 &$17.13\pm0.04$ &$17.75\pm0.05$ &$18.01\pm0.04$ &$18.44\pm0.11$&0.85&TI&0.130\\
97.10.18 &$17.20\pm0.01$ &$17.65\pm0.02$ &$17.97\pm0.02$ &$18.39\pm0.04$&0.74&TI&0.130\\
97.10.22 &$17.23\pm0.02$ &$17.73\pm0.05$ &$18.04\pm0.04$ &$18.50\pm0.12$&0.84&TI&0.130\\
97.10.23 &$17.18\pm0.02$ &$17.56\pm0.04$ &$17.96\pm0.04$ &$18.33\pm0.06$&0.83&TI&0.130\\
97.10.24 &$17.19\pm0.03$ &$17.76\pm0.04$ &$17.96\pm0.05$ &$18.54\pm0.07$&0.76&TI&0.130\\
97.11.11 &$17.24\pm0.02$ &$17.63\pm0.03$ &$17.95\pm0.05$ &$18.47\pm0.07$&0.77&TI&0.268\\
97.11.12 &$17.25\pm0.03$ &$17.70\pm0.05$ &$17.98\pm0.05$ &$18.46\pm0.08$&0.87&TI&0.268\\
98.07.23 &$17.10\pm0.04$ &$17.97\pm0.08$ &$17.52\pm0.10$ &$18.22\pm0.12$&0.96&TI&0.130\\ 
98.07.26 &$17.08\pm0.04$ &$17.82\pm0.04$ &$17.43\pm0.06$ &$18.12\pm0.07$&0.87&Pictor&0.159\\
98.07.28 &$17.14\pm0.02$ &$18.00\pm0.02$ &$17.63\pm0.06$ &$18.32\pm0.12$&0.88&TI&0.130\\
98.08.23 &$17.22\pm0.07$ &$18.03\pm0.04$ &$17.41\pm0.03$ &$18.33\pm0.06$&1.00&TI&0.130\\
98.08.24 &$17.18\pm0.02$ &$18.05\pm0.04$ &$17.39\pm0.03$ &$18.30\pm0.04$&1.05&TI&0.130\\
98.08.25 &$17.15\pm0.02$ &$17.97\pm0.02$ &$17.45\pm0.02$ &$18.28\pm0.04$&1.08&TI&0.130\\
98.08.26 &$17.15\pm0.03$ &$18.00\pm0.04$ &$17.45\pm0.03$ &$18.38\pm0.05$&0.89&TI&0.130\\
98.08.28 &$17.05\pm0.02$ &$17.85\pm0.02$ &$17.45\pm0.03$ &$18.04\pm0.04$&0.96&Pictor&0.159\\
98.08.29 &$17.06\pm0.02$ &$17.84\pm0.02$ &$17.39\pm0.03$ &$18.09\pm0.04$&0.86&Pictor&0.159\\
98.08.30 &$17.03\pm0.02$ &$17.82\pm0.02$ &$17.39\pm0.03$ &$17.99\pm0.04$&0.90&Pictor&0.159\\
98.08.31 &$17.06\pm0.02$ &$17.88\pm0.02$ &$17.40\pm0.03$ &$18.08\pm0.04$&0.86&Pictor&0.159\\
98.09.01 &$17.08\pm0.02$ &$17.83\pm0.02$ &$17.44\pm0.03$ &$18.04\pm0.04$&1.04&Pictor&0.159\\
98.09.02 &$17.06\pm0.02$ &$17.84\pm0.02$ &$17.40\pm0.03$ &$18.04\pm0.04$&0.86&Pictor&0.159\\
98.10.22 &$17.05\pm0.02$ &$17.89\pm0.02$ &$17.40\pm0.03$ &$17.99\pm0.04$&0.97&Pictor&0.159\\
98.11.14 &$17.07\pm0.05$ &$17.89\pm0.06$ &$17.43\pm0.05$ &$18.07\pm0.08$&1.05&TI&0.268\\
99.07.19 &$16.76\pm0.02$ &$18.01\pm0.04$ &$17.11\pm0.02$ &$18.08\pm0.03$&0.96&ST-7&0.160\\
99.07.20 &$16.75\pm0.06$ &$17.99\pm0.06$ &$17.14\pm0.04$ &$18.15\pm0.11$&1.15&ST-7&0.160\\
99.07.22 &$16.78\pm0.01$ &$18.00\pm0.03$ &$17.13\pm0.02$ &$18.09\pm0.02$&0.87&ST-7&0.160\\ 
\noalign{\smallskip} \hline
\end{tabular}
\end{center}
\end{table*}

\begin{table*}
\caption{Photometry of Q2237+0305 in $R$ band from observations in 2000 on Maidanak; 1.5-m
telescope, ST-7 CCD camera. Seing conditions (FWHM in arcseconds), and number of frames 
(n) are also presented. } 
\label{R2000}
\begin{center}
\begin{tabular}{ccccccc}
\hline
 \noalign{\smallskip}
 Date & A & B & C & D & FWHM & n \\
\hline   
00.07.23 &$16.729\pm0.004$ &$18.180\pm0.009$ &$17.806\pm0.016$ &$18.166\pm0.015$&1.08&16\\
00.07.24 &$16.726\pm0.005$ &$18.214\pm0.011$ &$17.826\pm0.016$ &$18.253\pm0.024$&1.36&28\\
00.07.25 &$16.731\pm0.003$ &$18.194\pm0.008$ &$17.794\pm0.007$ &$18.171\pm0.012$&1.11&30\\
00.07.26 &$16.734\pm0.003$ &$18.184\pm0.009$ &$17.802\pm0.006$ &$18.164\pm0.007$&0.78&8\\
00.07.28 &$16.743\pm0.005$ &$18.208\pm0.005$ &$17.808\pm0.007$ &$18.178\pm0.024$&0.89&7\\
00.07.29 &$16.722\pm0.003$ &$18.175\pm0.005$ &$17.799\pm0.005$ &$18.163\pm0.007$&0.94&35\\
00.07.30 &$16.738\pm0.003$ &$18.182\pm0.005$ &$17.790\pm0.004$ &$18.184\pm0.007$&0.86&20\\
00.08.01 &$16.728\pm0.010$ &$18.294\pm0.050$ &$17.887\pm0.038$ &$18.328\pm0.060$&1.42&7\\
00.08.02 &$16.731\pm0.003$ &$18.179\pm0.008$ &$17.788\pm0.007$ &$18.152\pm0.012$&0.77&12\\
00.08.03 &$16.740\pm0.004$ &$18.186\pm0.008$ &$17.819\pm0.010$ &$18.171\pm0.014$&1.14&21\\
00.08.04 &$16.731\pm0.004$ &$18.189\pm0.008$ &$17.803\pm0.008$ &$18.171\pm0.010$&0.88&14\\
00.08.05 &$16.745\pm0.002$ &$18.201\pm0.009$ &$17.832\pm0.008$ &$18.188\pm0.013$&1.09&34\\
00.08.06 &$16.749\pm0.005$ &$18.229\pm0.011$ &$17.834\pm0.007$ &$18.211\pm0.013$&1.16&35\\
00.08.07 &$16.733\pm0.008$ &$18.231\pm0.018$ &$17.867\pm0.021$ &$18.266\pm0.031$&1.46&21\\
00.08.08 &$16.726\pm0.002$ &$18.194\pm0.010$ &$17.821\pm0.008$ &$18.171\pm0.012$&1.07&23\\
00.08.09 &$16.736\pm0.004$ &$18.188\pm0.012$ &$17.803\pm0.009$ &$18.162\pm0.010$&0.89&9\\
00.08.10 &$16.726\pm0.004$ &$18.182\pm0.008$ &$17.789\pm0.010$ &$18.165\pm0.010$&0.89&9\\
00.08.11 &$16.737\pm0.003$ &$18.193\pm0.008$ &$17.807\pm0.012$ &$18.144\pm0.009$&0.99&10\\
00.08.12 &$16.725\pm0.004$ &$18.186\pm0.010$ &$17.800\pm0.009$ &$18.143\pm0.009$&0.98&17\\
00.08.13 &$16.736\pm0.002$ &$18.188\pm0.006$ &$17.811\pm0.005$ &$18.139\pm0.005$&0.96&42\\
00.08.18 &$16.723\pm0.006$ &$18.194\pm0.012$ &$17.807\pm0.013$ &$18.136\pm0.010$&0.92&10\\
00.08.19 &$16.741\pm0.004$ &$18.213\pm0.012$ &$17.857\pm0.010$ &$18.154\pm0.016$&1.04&15\\
00.08.20 &$16.734\pm0.003$ &$18.211\pm0.008$ &$17.809\pm0.006$ &$18.167\pm0.014$&1.13&32\\
00.08.22 &$16.739\pm0.004$ &$18.201\pm0.008$ &$17.825\pm0.008$ &$18.158\pm0.012$&0.93&12\\
00.08.23 &$16.744\pm0.005$ &$18.209\pm0.009$ &$17.820\pm0.008$ &$18.165\pm0.013$&0.86&10\\
00.08.24 &$16.732\pm0.003$ &$18.208\pm0.008$ &$17.828\pm0.007$ &$18.167\pm0.011$&0.92&20\\
00.08.25 &$16.743\pm0.005$ &$18.188\pm0.007$ &$17.825\pm0.009$ &$18.135\pm0.007$&0.85&10\\
00.08.26 &$16.742\pm0.008$ &$18.221\pm0.013$ &$17.831\pm0.015$ &$18.176\pm0.016$&1.02&15\\
00.08.30 &$16.747\pm0.006$ &$18.212\pm0.010$ &$17.821\pm0.007$ &$18.168\pm0.017$&1.12&17\\
00.08.31 &$16.736\pm0.004$ &$18.214\pm0.007$ &$17.819\pm0.010$ &$18.161\pm0.013$&0.96&12\\
00.09.01 &$16.737\pm0.004$ &$18.190\pm0.009$ &$17.815\pm0.009$ &$18.144\pm0.013$&0.85&20\\
00.09.02 &$16.752\pm0.006$ &$18.201\pm0.012$ &$17.831\pm0.010$ &$18.144\pm0.010$&1.00&10\\
00.09.04 &$16.749\pm0.005$ &$18.226\pm0.017$ &$17.841\pm0.020$ &$18.164\pm0.019$&0.92&9\\
00.09.05 &$16.755\pm0.003$ &$18.220\pm0.010$ &$17.859\pm0.010$ &$18.152\pm0.013$&1.16&21\\
00.09.06 &$16.757\pm0.003$ &$18.245\pm0.011$ &$17.863\pm0.012$ &$18.162\pm0.012$&1.13&12\\
00.09.07 &$16.746\pm0.005$ &$18.207\pm0.006$ &$17.833\pm0.006$ &$18.120\pm0.008$&0.80&10\\
00.09.08 &$16.743\pm0.004$ &$18.204\pm0.010$ &$17.837\pm0.011$ &$18.140\pm0.012$&1.13&22\\
00.09.09 &$16.744\pm0.004$ &$18.223\pm0.012$ &$17.834\pm0.008$ &$18.142\pm0.021$&0.97&10\\
00.09.10 &$16.762\pm0.003$ &$18.235\pm0.026$ &$17.864\pm0.013$ &$18.175\pm0.012$&0.96&10\\
00.09.23 &$16.762\pm0.005$ &$18.238\pm0.018$ &$17.849\pm0.012$ &$18.095\pm0.027$&1.04&11\\
00.09.24 &$16.746\pm0.010$ &$18.264\pm0.028$ &$17.851\pm0.026$ &$18.151\pm0.027$&1.06&10\\
00.09.26 &$16.772\pm0.005$ &$18.253\pm0.017$ &$17.855\pm0.009$ &$18.118\pm0.018$&1.16&10\\
00.09.27 &$16.749\pm0.005$ &$18.240\pm0.008$ &$17.845\pm0.008$ &$18.111\pm0.013$&1.18&20\\
00.09.28 &$16.764\pm0.005$ &$18.273\pm0.015$ &$17.845\pm0.010$ &$18.143\pm0.017$&1.04&10\\
00.09.30 &$16.784\pm0.008$ &$18.265\pm0.019$ &$17.909\pm0.020$ &$18.149\pm0.021$&1.21&10\\
00.10.07 &$16.776\pm0.012$ &$18.294\pm0.037$ &$18.001\pm0.034$ &$18.075\pm0.033$&1.52&11\\
\noalign{\smallskip} \hline
\end{tabular}
\end{center}
\end{table*} 

\begin{table*}
\caption{$VRI$ Photometry of Q2237+0305 in 1995-2000; Maidanak, 1.5-m telescope.} 
\label{vri95-00}
\begin{center}
\begin{tabular}{cccccc}
\hline
 \noalign{\smallskip}
Date & A & B & C & D & Sp.band \\
\hline   
         &$17.34\pm0.04$  &$17.44\pm0.03$  &$18.41\pm0.10$  &$18.66\pm0.08$&V\\
95.09.17 &$17.18\pm0.03$  &$17.32\pm0.03$  &$18.13\pm0.06$  &$18.44\pm0.07$&R\\
         &$16.99\pm0.03$  &$17.21\pm0.04$  &$17.83\pm0.08$  &$18.23\pm0.09$&I\\
\hline          
         &$17.43\pm0.03$  &$17.95\pm0.05$  &$18.42\pm0.05$  &$18.78\pm0.07$&V\\
97.08.29 &$17.15\pm0.03$  &$17.72\pm0.03$  &$18.08\pm0.12$  &$18.38\pm0.10$&R\\  
         &$16.96\pm0.03$  &$17.60\pm0.04$  &$17.81\pm0.04$  &$18.15\pm0.07$&I\\
\hline           
         &$17.40\pm0.03$  &$17.94\pm0.04$  &$18.35\pm0.05$  &$18.79\pm0.06$&V\\
97.08.30 &$17.14\pm0.02$  &$17.75\pm0.03$  &$18.08\pm0.04$  &$18.41\pm0.05$&R\\     
         &$16.89\pm0.02$  &$17.55\pm0.04$  &$17.82\pm0.03$  &$18.11\pm0.06$&I\\
\hline  
         &$17.47\pm0.05$  &$18.01\pm0.07$  &$18.38\pm0.06$  &$18.91\pm0.09$&V\\
97.09.01 &$17.13\pm0.04$  &$17.75\pm0.05$  &$18.01\pm0.04$  &$18.44\pm0.11$&R\\
         &$16.93\pm0.03$  &$17.61\pm0.04$  &$17.83\pm0.04$  &$18.19\pm0.09$&I\\
\hline                  
         &$17.30\pm0.02$  &$18.11\pm0.05$  &$17.66\pm0.05$  &$18.52\pm0.05$&V\\
98.07.26 &$17.08\pm0.04$  &$17.82\pm0.04$  &$17.43\pm0.06$  &$18.12\pm0.07$&R\\   
         &$16.92\pm0.03$  &$17.71\pm0.04$  &$17.32\pm0.04$  &$17.87\pm0.06$&I\\  
\hline       
         &$17.44\pm0.03$  &$18.23\pm0.03$  &$17.78\pm0.03$  &$18.64\pm0.05$&V\\ 
98.07.28 &$17.14\pm0.02$  &$18.00\pm0.02$  &$17.63\pm0.06$  &$18.32\pm0.12$&R\\  
         &$16.93\pm0.04$  &$17.72\pm0.05$  &$17.47\pm0.05$  &$17.82\pm0.06$&I\\
\hline                 
         &$17.52\pm0.02$  &$18.23\pm0.02$  &$17.77\pm0.04$  &$18.84\pm0.07$&V\\
98.08.23 &$17.22\pm0.07$  &$18.03\pm0.04$  &$17.41\pm0.03$  &$18.33\pm0.06$&R\\
         &$17.09\pm0.04$  &$17.83\pm0.03$  &$17.36\pm0.04$  &$18.14\pm0.20$&I\\   
\hline          
         &$17.34\pm0.05$  &$18.28\pm0.05$  &$17.65\pm0.06$  &$18.48\pm0.11$&V\\
98.11.14 &$17.07\pm0.05$  &$17.89\pm0.06$  &$17.43\pm0.05$  &$18.07\pm0.08$&R\\
         &$16.94\pm0.03$  &$17.79\pm0.05$  &$17.44\pm0.03$  &$17.92\pm0.06$&I\\
\hline 
         &$16.89\pm0.04$  &$18.12\pm0.05$  &$17.19\pm0.03$  &$18.34\pm0.11$&V\\
99.07.20 &$16.75\pm0.06$  &$17.99\pm0.06$  &$17.14\pm0.04$  &$18.15\pm0.11$&R\\
         &$16.62\pm0.02$  &$17.76\pm0.05$  &$17.07\pm0.02$  &$17.84\pm0.04$&I\\
\hline
         &$16.92\pm0.02$  &$18.16\pm0.03$  &$17.22\pm0.02$  &$18.40\pm0.08$&V\\
99.07.22 &$16.78\pm0.01$  &$18.00\pm0.03$  &$17.13\pm0.02$  &$18.09\pm0.02$&R\\ 
         &$16.63\pm0.01$  &$17.79\pm0.03$  &$17.09\pm0.02$  &$17.82\pm0.03$&I\\
\hline         
         &$16.888\pm0.005$&$18.401\pm0.016$&$18.011\pm0.008$&$18.400\pm0.007$&V\\
00.07.26 &$16.734\pm0.003$&$18.184\pm0.009$&$17.802\pm0.006$&$18.164\pm0.007$&R\\
         &$16.577\pm0.004$&$17.973\pm0.007$&$17.590\pm0.008$&$17.911\pm0.006$&I\\
\hline
         &$16.880\pm0.004$&$18.387\pm0.015$&$17.992\pm0.011$&$18.392\pm0.017$&V\\
00.08.04 &$16.731\pm0.004$&$18.189\pm0.008$&$17.803\pm0.008$&$18.171\pm0.010$&R\\
         &$16.577\pm0.004$&$17.968\pm0.007$&$17.604\pm0.006$&$17.919\pm0.004$&I\\
\hline        
         &$16.894\pm0.008$&$18.423\pm0.010$&$18.032\pm0.017$&$18.381\pm0.020$&V\\
00.08.09 &$16.736\pm0.004$&$18.188\pm0.012$&$17.803\pm0.009$&$18.162\pm0.010$&R\\  
         &$16.586\pm0.005$&$17.976\pm0.008$&$17.616\pm0.010$&$17.925\pm0.016$&I\\
\hline
         &$16.878\pm0.008$&$18.365\pm0.025$&$18.021\pm0.030$&$18.350\pm0.025$&V\\
00.08.18 &$16.723\pm0.006$&$18.194\pm0.012$&$17.807\pm0.013$&$18.136\pm0.010$&R\\
         &$16.574\pm0.006$&$17.975\pm0.010$&$17.612\pm0.013$&$17.899\pm0.012$&I\\
\hline          
         &$16.882\pm0.006$&$18.369\pm0.014$&$18.008\pm0.010$&$18.370\pm0.015$&V\\
00.08.25 &$16.743\pm0.005$&$18.188\pm0.007$&$17.825\pm0.009$&$18.135\pm0.007$&R\\
         &$16.578\pm0.003$&$17.982\pm0.008$&$17.616\pm0.004$&$17.884\pm0.011$&I\\
\hline
         &$16.886\pm0.006$&$18.398\pm0.013$&$17.997\pm0.007$&$18.328\pm0.016$&V\\
00.09.07 &$16.746\pm0.005$&$18.207\pm0.006$&$17.833\pm0.006$&$18.120\pm0.008$&R\\
         &$16.583\pm0.006$&$17.988\pm0.010$&$17.612\pm0.008$&$17.884\pm0.012$&I\\
\hline           
         &$16.918\pm0.005$&$18.443\pm0.011$&$18.057\pm0.013$&$18.365\pm0.016$&V\\
00.09.28 &$16.764\pm0.005$&$18.273\pm0.015$&$17.845\pm0.010$&$18.143\pm0.017$&R\\
         &$16.600\pm0.006$&$18.006\pm0.013$&$17.657\pm0.006$&$17.871\pm0.016$&I\\       
\noalign{\smallskip} \hline
\end{tabular}
\end{center}
\end{table*}

\begin{table*}
\begin{center}
\caption{$(V-R)$ and $(R-I)$ colors of Q2237+0305 A,B,C,D; 1995-2000, Maidanak, 
1.5-m
telescope.}
\label{colors}
\begin{tabular}{c|cc|cc|cc|cc}
\hline
 \noalign{\smallskip}

Date &\multicolumn{2}{c}{A}&\multicolumn{2}{c}{B}&\multicolumn{2}{c}{C}
     &\multicolumn{2}{c}{D}\\
\cline{2-9}
\noalign{\smallskip}
         &  V-R  &   V-I &  V-R  &  V-I  &  V-R  &  V-I  &  V-R  &  V-I\\
\hline \noalign{\smallskip}
17.09.95 &  0.16 &  0.35 &  0.12 &  0.23 &  0.28 &  0.42 &  0.22 &  0.57\\
28.08.97 &  0.28 &  0.47 &  0.23 &  0.35 &  0.34 &  0.51 &  0.40 &  0.63\\
30.08.97 &  0.26 &  0.51 &  0.19 &  0.39 &  0.27 &  0.63 &  0.38 &  0.68\\
01.09.97 &  0.34 &  0.54 &  0.26 &  0.40 &  0.37 &  0.55 &  0.47 &  0.72\\
26.07.98 &  0.22 &  0.38 &  0.29 &  0.40 &  0.23 &  0.34 &  0.40 &  0.65\\
28.07.98 &  0.30 &  0.51 &  0.23 &  0.51 &  0.15 &  0.31 &  0.32 &  0.70\\
23.08.98 &  0.30 &  0.43 &  0.20 &  0.60 &  0.36 &  0.41 &  0.51 &  0.70\\
14.11.98 &  0.27 &  0.40 &  0.39 &  0.59 &  0.22 &  0.21 &  0.41 &  0.56\\
20.07.99 &  0.14 &  0.27 &  0.13 &  0.34 &  0.05 &  0.12 &  0.19 &  0.50\\
22.07.99 &  0.14 &  0.29 &  0.16 &  0.37 &  0.09 &  0.13 &  0.31 &  0.58\\
26.07.00 &  0.15 &  0.31 &  0.22 &  0.43 &  0.21 &  0.42 &  0.24 &  0.49\\
04.08.00 &  0.15 &  0.30 &  0.20 &  0.42 &  0.19 &  0.39 &  0.22 &  0.47\\
09.08.00 &  0.15 &  0.31 &  0.23 &  0.45 &  0.23 &  0.42 &  0.22 &  0.46\\
18.08.00 &  0.15 &  0.30 &  0.17 &  0.39 &  0.21 &  0.41 &  0.21 &  0.45\\
25.08.00 &  0.14 &  0.30 &  0.18 &  0.39 &  0.18 &  0.39 &  0.23 &  0.49\\
07.09.00 &  0.14 &  0.30 &  0.19 &  0.41 &  0.16 &  0.38 &  0.22 &  0.44\\
28.09.00 &  0.15 &  0.32 &  0.17 &  0.44 &  0.21 &  0.40 &  0.22 &  0.39\\
\hline \noalign{\smallskip}
\end{tabular}
\end{center}
\end{table*}

\centerline{\bf Figure captions}\vspace{0.3cm}
\noindent Fig. 1. Images of Q2237+0305 for six epochs, obtained in $R$ band 
with the 1.5-m Maidanak telescope; A component is at the bottom, C is at the left.

\vspace{0.2cm}
\noindent Fig. 2. Photometry of Q2237+0305 A,B,C,D in $R$ band from observations
 with the 1.5-m Maidanak telescope in 1997-2000, (large symbols). Photometry 
 in $V$ band by OGLE program is also plotted by smaller and fainter symbols. 
 Our data (Tables 4 and 5) are shifted arbitrarily for better comparison, 
 see Sec. 4 for more details. The apparent brightness discrepancy near the 
 image C 1999 brightness peak results from the different monitoring filter 
 bands used and a significant color change during the brightness peak event.
  
\vspace{0.2cm}
\noindent Fig.3. Photometry of Q2237+0305 A,B,C,D in $R$ band from observations
 with the 1.5-m Maidanak telescope in 2000, July 23 - September 7. The light 
 curves of the  components are shifted arbitrarily along the magnitude axis 
 with respect to each other  for clarity, and approximated by second-order 
 polynomials.

\vspace{0.2cm}
\noindent Fig. 4. $(V-I)$ colors vrs $R$ magnitudes, calculated for observations of 
 1995-2000. Note a separate cluster of points for A component. A regression 
 line slope for B, C and D components is of $0.33 \pm 0.08$.

\vspace{0.2cm}
\noindent Fig. 5. Variations of $(V-I)$ colors (vertical axis) vrs variations of $R$ 
 magnitudes, calculated for observations of 1995-2000. A regression line 
 slope is $0.31\pm 0.08.$

\vspace{0.2cm}
\noindent Fig. 6. Color $(V-I)$ vrs color $(V-R)$ diagram for Q2237+0305 A,B,C,D
 components, plotted from the Maidanak data, taken in 1995-2000. A regression 
 line slope is $1.31 \pm 0.14.$
 
 \vspace{0.2cm}
\noindent Fig. 7. Long-term variations of $R$ magnitudes (upper panel) and $(V-I)$ 
 colors (at the bottom) of A, B, C, D components of Q2237+0305 from the 
 observations of 1995-2000. Each point is a result of averaging within 
 one observational set.


\vspace{1cm}
Figures 1--7 " are available in "gif" format from:\\
\vspace{1cm}
\centerline
{http://arXiv.org/ps/astro-ph/}

\end{document}